# Anomalous Gulf Heating and Hurricane Katrina's Rapid Intensification

Menas Kafatos, Ritesh Gautam, Guido Cervone, Zafer Boybeyi & Donglian Sun

Center for Earth Observing and Space Research
School of Computational Sciences
George Mason University
Fairfax, VA 22030, USA

Global warming due to the increasing concentration of greenhouse gases has become a great concern and has been linked to increased hurricane activity (*1-3*) associated with higher sea surface temperatures (SSTs) with conflicting views (*4*). Our observational results based on long term trends of SST reveal that the anomaly reached a record 0.8 °C in the Gulf of Mexico in August 2005 as compared to previous years and may have been responsible for the intensification of the devastating Hurricane Katrina into a category 5 hurricane that hit the Southern coast of United States severely impacting the low lying city of New Orleans and the surrounding areas. In most intensifying storms, air-sea interaction is the major contributing factor (*5*) and here we show how air-sea interactions might have affected Katrina's rapid intensification in the Gulf.

Air-sea interactions include role of latent heat flux (*6*), sensible heat flux and SST towards the intensification and decay of hurricanes. Generally, this requires the SST to be at least 26 °C for hurricanes to form, be maintained and intensify (*7*). Over warm waters, latent heat releases moisten the lower troposphere, while sensible heat fluxes warm the lower troposphere. Additionally, within the hurricane, there is a release of latent heat due to water vapor cooling and condensing into cloud droplets. This causes heat energy to be converted from potential energy to kinetic energy thus producing a secondary circulation within the hurricane. The overall effect is the feedback and intensification of the process.

Figure 1A shows that the mean SST values were above ~30 °C with a discernible warm patch of ~33 °C in the northwestern shelf adjacent to the state of Louisiana in August, 2005. The SST anomaly (SSTA) reached a record value of ~0.8 °C as compared to previous years (Fig. 1B) with a positively increasing trend. The northwestern shelf in the upper Gulf can be identified as a localized warm area that might have provided additional fuel to the hurricane as it entered the Gulf from the western side crossing southern Florida. Katrina was category 1 on August 25 and intensified to Category 2, 3, 4 and finally 5 after entering the Gulf. Furthermore, daily variations of surface latent heat flux (SLHF) and sensible heat flux (SHF) values show significant increases during the intensification period of the hurricane with large peak values of ~500 $Wm^{-2}$ and ~70 $Wm^{-2}$, respectively (Fig. 1C). We estimated that the heat contained in the abnormally elevated SST values of the Gulf is comparable to the energy in latent heat in the inner hurricane, ~ $10^{14}$ J.

These results suggest that the anomalous SST was the prominent factor causing the intensification of Katrina in the Gulf, although other associated parameters such as vertical wind shear of ambient flow and atmospheric stability were also important. The increasing trend of SSTA from 2001 combined with the increased hurricane activities in recent years suggest that the possibility of climate anomalies induced by global warming may have played a vital role in the intensification of Katrina. Tracking anomalously warm Gulf waters is important since if this trend continues in the years ahead, an increased number of hurricanes like Katrina will continue striking the Gulf areas.

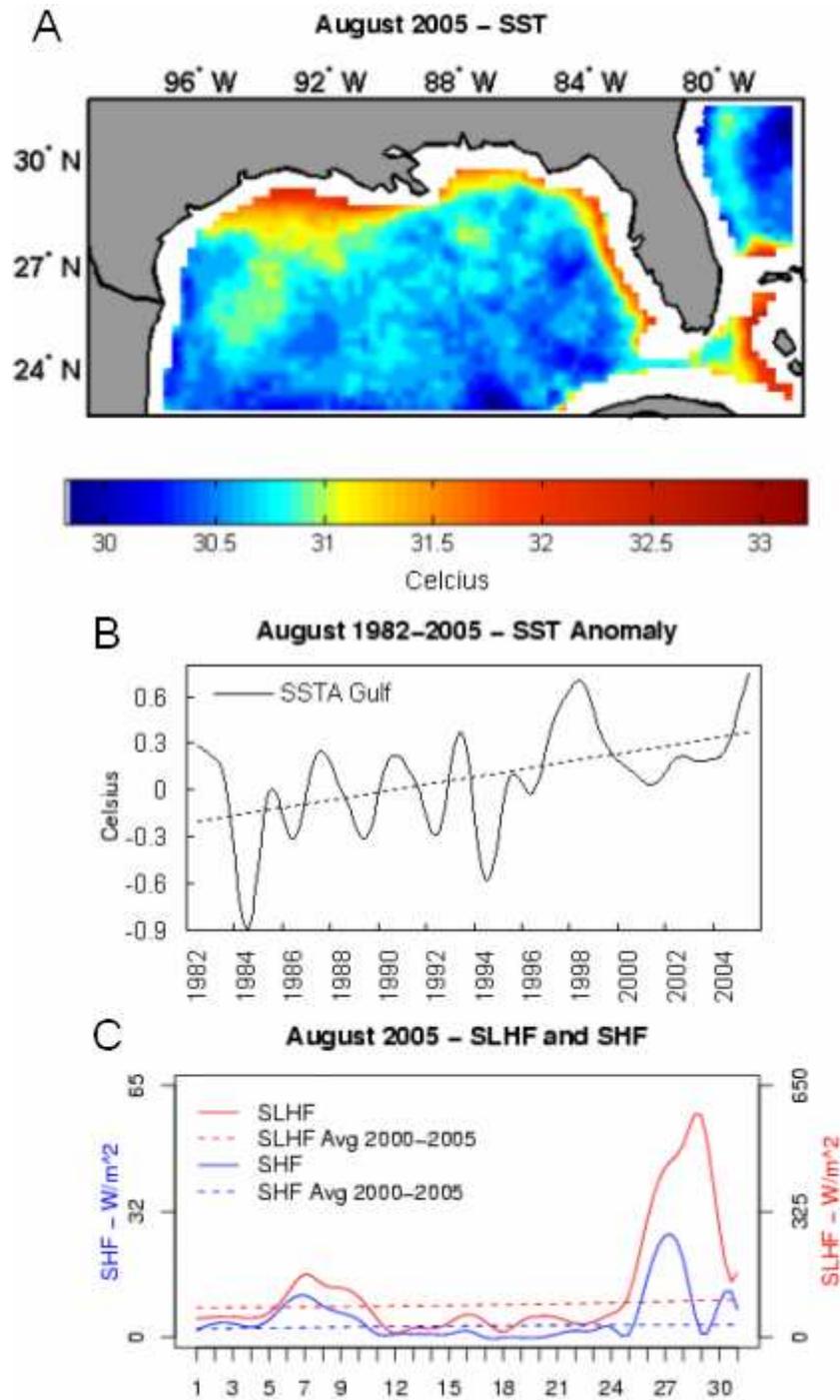

**Fig. 1. (A)** Spatial distribution of SST in August 2005 in the Gulf of Mexico. **(B)** SSTA for August from 1982-2005 peaking in August 2005 with an increasing positive trend (the trend is indicated by dashed line). **(C)** Daily variations of SLHF and SHF values for August 2005 (their average values for the period 2000-2005 are indicated by dashed lines).